\documentclass[conference,final,twocolumn]{IEEEtran}
\usepackage{amsmath,amssymb,amsfonts,mathrsfs,bm}
\usepackage{amstext}
\usepackage{upgreek}
\usepackage{multicol}
\usepackage{indentfirst}
\usepackage{graphicx}
\usepackage{paralist}
\usepackage{cite}
\usepackage{citesort}
\usepackage{booktabs}
\usepackage{multirow}

\IEEEoverridecommandlockouts

\begin{document}
\title{Cooperative Device-to-Device Communications With Caching}
\author{

\IEEEauthorblockN{{Binqiang Chen and Chenyang Yang}} \vspace{0.2cm}
\IEEEauthorblockA{Beihang University, Beijing, China\\
Email: \{chenbq,cyyang\}@buaa.edu.cn}
\and \IEEEauthorblockN{{Gang Wang}} \vspace{0.2cm}
\IEEEauthorblockA{NEC Labs, China \\
Email: wang\_gang@nec.cn}
\thanks{This work was supported by China NSFC under Grant 61429101 and NEC Labs.}
%
}

\maketitle

\begin{abstract}
Device-to-Device (D2D) communications can increase the throughput of cellular networks significantly, where the interference among D2D links should be properly managed. In this paper, we propose an opportunistic cooperative D2D transmission
strategy by exploiting the caching capability at the users to deal with the interference among D2D links. To increase the cooperative opportunity and improve spatial reuse gain,
we divide the D2D users into clusters and cache different popular files at the users within a cluster, and then find the optimal cluster size. To maximize the
network throughput, we assign different frequency bands to cooperative and non-cooperative D2D links and optimize the bandwidth
partition. Simulation results demonstrate that the proposed strategy can provide $500\% \sim 600\%$ throughput gain over existing cache-enabled D2D
communications when the popularity distribution is skewed, and can provide $40\% \sim 80\%$ gain even when the popularity distribution is uniform.
\end{abstract}

\begin{keywords}
Caching, D2D, Cooperative transmission
\end{keywords}\vspace{-0.01mm}

\section{Introduction}

Device-to-device (D2D) communications is a promising technique to boost the throughput for fifth-generation (5G) cellular networks \cite{Andrews.D2D}. Its typical use-cases include
cellular offloading, content distribution, and relaying, \emph{etc.} \cite{Survey.D2D}, where content delivery has attracted considerable attention recently.

Motivated by the observation that a large amount of traffic is generated by a few contents while the storage of mobile devices grows rapidly with low cost,
the authors of \cite{Golrezaei.TWC,JMY.JSAC} proposed to cache popular files on mobile devices and then employ D2D communications
to increase the  network throughput. Nonetheless, the interference between active D2D links is simply treated as noise \cite{Golrezaei.TWC} or managed using time division multiple
access (TDMA) \cite{JMY.JSAC}. This inevitably limits the throughput gain.

In \cite{IA.D2D}, the authors proposed to apply interference alignment (IA) to  mitigate the interference  among D2D links, However, only
three D2D links are coordinated within each cluster, and the interference among the clusters are treated as noise.
In \cite{CD2D.15,GC13.Relay,C2D.Relay}, cooperative relay techniques were proposed to mitigate the interference between cellular
and D2D links, which can not manage the interference among D2D links.

In fact, when contents are cached at transmitters, cooperative transmission becomes possible, which is a more effective way to deal with interference than IA. In \cite{Lau.Tran13}, the authors proposed a base station (BS) cooperative transmission strategy by exploiting caches at BSs, where precoding and caching policy are jointly optimized. In the network with D2D links, if some D2D transmitters (DTs) have cached the files requested by some D2D receivers (DRs), these DTs can jointly transmit the requested files
to these DRs without the need of exchanging data.
This  strategy of \emph{opportunistic cooperative D2D communications with caching} is called as \emph{Coop strategy} in the sequel for
simplicity.

This strategy is viable in practice. First, in D2D communications, a DT can assist other users in additional to transmitting data to its
destined DR, e.g., cooperative relay \cite{GC13.Relay}. Second, the channel state information among D2D links required by the \emph{Coop} strategy can be obtained at DTs
and the BS through channel probing and feedback \cite{CSI}, while the synchronization among cooperated DTs can be realized with the assistance of
the BS \cite{Andrews.D2D} or can be realized at users with the methods proposed in \cite{SYN.13}.



In this paper, we propose a  Coop strategy to manage the interference among D2D links.  To facilitate cooperative transmission and exploit spatial reuse, we divide a cell into virtual clusters and cache different popular files at the users within each cluster. To avoid mutual interference, we assign different frequency bands to cooperative and non-cooperative D2D links. Aimed at improving the network
throughput while ensure user fairness, we first optimize  cluster size to maximize the average number of cooperative users, and then optimize bandwidth partition to maximize the average network throughput under the constraint on average user rate.


\section{System Model}
\label{sec:system model} Consider a cellular network, where $M$ single-antenna users are uniformly located in a hotpot area within a macro cell,
which is assumed as a square area with side length of $D_c$. The square hotpot area is divided into $B$ smaller square areas called clusters as in
\cite{Golrezaei.TWC}, where the side length of each  cluster is $D=D_c/\sqrt{B}$. For mathematical simplicity, we assume that the number of users per
cluster is $K=M/B$ and each user transmits with its maximal power $P$ as in \cite{JMY.JSAC}.

Each user can cache $N$ files. The BS is
aware of the cached files of the users and coordinates the D2D communications.
We consider static content catalog including $N^f$ files that the users may request, where the files are indexed according to the popularity, e.g., the 1st file
is the most popular file. According to the user cache capacity $N$, all files are divided into $K_0=N^f/N$ groups, where  the $k$th \emph{file group} $\mathcal{G}_k$ consists of the  files with indices $(k-1)N+1, \cdots, kN$, $1 \leq k\leq K_0$, e.g., the 1st file group $\mathcal{G}_1$ contains the most popular
$N$ files.

Because the users usually do not allow the operator to occupy a large portion of their storage space, not all the files possibly interested by the users in a cell can be cached at the users. Hence, $K \leq K_0$.

The probability that the $i$th file is requested by a user is assumed to follow Zipf distribution, which is $P_{N^f}(i)=i^{-\beta}/\sum_{k=1}^{N^f}k^{-\beta}$,
where the parameter $\beta$ reflects the popularity of the files \cite{Zipf99}. Then, the probability that a user requests a file within the $k$th \emph{file group} $\mathcal{G}_k$ can be obtained as
\begin{equation}
\label{equ.P_k}
P_k=\frac{\sum_{j=(k-1)N+1}^{kN}j^{-\beta}}{\sum_{i=1}^{N^f}i^{-\beta}} .
\end{equation}

To increase the cooperative opportunity and improve spatial reuse, we consider the following cache placement policy as in \cite{Golrezaei.TWC}. In every cluster, the $k$th user caches the $k$th file
group $\mathcal{G}_k$. Since each cluster contains $K$ users, the file groups with indices exceeding $K$ are not cached at users. Hence,
the most popular $KN$ files are cached in every cluster. We consider such a policy since only when the users within a cluster cache different
files, the probability that one user can fetch files through D2D links can be maximized. In practice, these files can be proactively downloaded by the operator
from the  BS to the caches at each user via broadcasting during off-peak time according to the  user demand statistics.

Considering that the users within each cluster cannot cooperate due to caching different files, we randomly select one D2D link in each cluster to transmit in the same time-frequency resource to avoid intra-cluster interference.

If the file requested by a user is cached at any other user in the cluster it belongs to (called \emph{local cluster of
the user}), then the user can directly obtain the file with D2D communication. If the file requested by a user is in its local cache, it
can fetch the file immediately with zero delay, but we ignore this case for analysis simplicity as in
\cite{JMY.JSAC}. Otherwise, the file will be fetched via backhaul and then transmitted to the user by the BS. We
consider overlay inband D2D \cite{Survey.D2D}, and assume that a fixed bandwidth of $W$ is assigned to the D2D links.

Then, all the $M$ users can be classified into three types:
\begin{itemize}
\item Cellular users: These users need to fetch their requested files from the BS, whose number is denoted as $N^b$.
\item Coop users: If there exist users in a cluster requesting the files in $\mathcal{G}_k$, then the cluster \emph{hits the $k$th file group}. If every cluster hits the same file group $\mathcal{G}_k$, the users requesting the files in $\mathcal{G}_k$ can establish cooperative D2D
    links, where the $k$th user  in each cluster who caches $\mathcal{G}_k$ acting as DT, and the DTs can jointly transmit to the users who request the files in $\mathcal{G}_k$ (called Coop D2D users or Coop users for simplicity), whose number is denoted as $N^c$.
\item N-Coop users: The remaining users except the cellular and Coop users are Non-Coop D2D users (N-Coop users for simplicity), whose number is $N^n= M-N^b-N^c$.
\end{itemize}

To avoid mutual interference, we assign $\eta W$ for Coop users and remaining bandwidth $(1-\eta) W$ for N-Coop users, where $\eta$ is the bandwidth allocation factor and $0 \leq \eta < 1$.

This is an opportunistic Coop strategy, which may operate in the following two modes. In \emph{Mode} $0$, there exist clusters not hitting the file group $\mathcal{G}_k$ for any $k$. Then, all the DTs transmit independently, and hence all the bandwidth is assigned to the N-Coop users, i.e.,
$\eta=0$. In \emph{Mode} $1$, there exist file groups hit by every cluster. Then, there exist both Coop and N-Coop users, i.e., $0<\eta<1$.


\section{Optimizing Cluster Size}
\label{sec:cluster}
In this section we optimize the cluster size characterized by the number of users per cluster to maximize
the average active Coop users.
Since only one D2D link is active in each cluster each time, $B$ users out of all Coop users can be scheduled simultaneously in \emph{Mode} 1. Therefore, the
number of active Coop users $N^a=B$ in \emph{Mode} $1$, and $N^a=0$ in \emph{Mode} 0. $N^a$ reflects how many interference-free D2D links can be active concurrently. A large value of $N^a$ indicates a high throughput. Note that $N^a \ne N^c$.

A cluster hits the $k$th file group if at least one of the $K$ users in the cluster requests a file in $\mathcal{G}_k$, whose probability  can be obtained from
(\ref{equ.P_k})  as
\begin{equation}
\label{equ.P^h_k}
P^h_k=1-(1-P_k)^K,
\end{equation}
which increases with  $K$.
Then, the probability that there exist file groups hit by all the $B$ clusters (i.e., the
network operates in Mode 1, called cooperative probability) can be derived
as
\begin{equation}
\label{equ.P^c}
P^c=1-\prod_{k=1}^{K}(1-(P^h_k)^B),
\end{equation}
where $1-(P^h_k)^B$  decreases with
$K$ since $B=M/K$. Hence, $P^c$ is an increasing function of $K$.

Consequently, the average active Coop users is
\begin{equation}
\label{equ.E_N^a}
\mathbb{E}\{N^a\}= BP^c+0(1-P^c)=BP^c.
\end{equation}
When the number of users per cluster $K$ is large, the cooperative probability is high, but the number of active Coop users  $N^a=M/K$ is small. This suggests that there is a
tradeoff between two counter-running effects: a small value of $K$ means more active Coop users in Mode 1, and a large value of $K$ means high cooperative probability.

Thus, the optimal cluster size that maximizes $\mathbb{E}\{N^a\}$ can be found from the following problem
\begin{equation}
\label{equ.opt1}
\begin{aligned}
&\max_K \,\, BP^c\\
&s.t.\quad
BK=M, \quad
1\leq K \leq K_0.
\end{aligned}
\end{equation}

The optimal number of users per cluster $K^*$ can be found by one-dimensional searching, which is of low complexity.
From \eqref{equ.P_k}, \eqref{equ.P^h_k} and \eqref{equ.P^c}, we can see that $K^*$ depends on the  catalog size $N_f$, the popularity parameter $\beta$, the number of users in the hotpot area $M$ and the user cache capacity $N$, which do not often change. Hence, it is unnecessary to update the optimal cluster frequently.

\section{Optimizing Bandwidth Partition}
\label{sec:allocation} Since the numbers of Coop users and N-Coop users are random and hardly equal, the spectral
efficiency will be reduced if we simply assign identical bandwidth to these two types of users.
In this section, we optimize the bandwidth partition to maximize the average throughput of the  network  under the fairness constraints of the users.

\subsection{Average Throughput}
Since only one D2D link is active in a cluster in each time, 
the average throughput of the network operating in \emph{Mode} 0 can be obtained as follows,
\begin{equation}
\label{equ.bar_R_0}
\bar{R_0} = \mathbb{E}\{W\sum_{i=1}^{B}R_i^n\} \overset{(b)}{=} WB\bar{R_i^n},
\end{equation}
where ${R_i^n}$ and $\bar{R_i^n}$ are respectively the instantaneous and average throughputs of a N-Coop link per unit bandwidth, and the expectation is taken over small scale channel fading and user location. Since all users are randomly located and transmit with equal power, $(b)$ can be obtained.

Analogically, the average throughput of the network operating in \emph{Mode} 1 can be obtained as
\begin{equation}
\label{equ.bar_R_1}
\begin{split}
&\bar{R_1} = \mathbb{E}\{ \eta W\sum_{i=1}^{B}R_i^c\ + (1-\eta)W\sum_{i=1}^{B}R_i^n\}\\
& = WB(\eta \bar{R_i^c}+(1-\eta) \bar{R_i^n} ),
\end{split}
\end{equation}
where ${R_i^c}$ and $\bar{R_i^c}$ are respectively the instantaneous and average throughputs per unit bandwidth of a Coop link.

Further considering the cooperative probability in (\ref{equ.P^c}), the average throughput of the network is
\begin{equation}
\label{equ.bar_R}
\bar{R} = P^c\bar{R_1} + (1-P^c)\bar{R_0} = WB(P^c\eta \bar{R_i^c}+(1-P^c\eta)\bar{R_i^n}).
\end{equation}

\subsubsection{Average Throughput of N-Coop Link}

Without cooperation, a DT delivers the requested file to its corresponding DR in a way by treating the inter-cluster interference as
noise. Then, the signal to interference plus noise ratio (SINR) at the DR of the active link in the $i$th cluster can be
expressed as
\begin{equation}
\label{equ.SINR_R^n}
\gamma_i^n = \frac{P|h_{ii}|^2r_{ii}^{-\alpha}}{I_i+\sigma^2},
\end{equation}
where $P$ is the transmit power, $\sigma^2$ is the variance of white Gaussian noise, $I_i = P\sum_{j=1,j \neq i}^{B}r_{ij}^{-\alpha}|h_{ij}|^2$
is the power of inter-cluster interference, $h_{ij}$ and $r_{ij}$ are respectively the channel coefficient and distance between the DT  and the DR with $h_{ij}$ following complex Gaussian distribution with zero mean and unit variance, and  $\alpha$ is the path loss exponent.

Due to the short distance of a  D2D link, we consider interference-limited scenario and assume $I_i \gg \sigma^2$. Then,
the N-Coop link throughput in unit bandwidth is
$R_i^n = \log_2(1+\frac{P|h_{ii}|^2r_{ii}^{-\alpha}}{I_i})$.
Considering that $|h_{ij}|^2$ follows Exponential distribution, which is a special case of Gamma
distribution, the interference power $I_i$ can be approximated as Gamma distribution
\cite{Gamma.2011}. Further considering that for Gamma distributed random variable $X$ with parameters $k$ and $\theta$,
$\mathbb{E}\{\ln(X)\}=\psi(k)+\ln(\theta)$, where $\psi(k)$ is the Digamma function, the average  throughput of a  N-Coop link per unit bandwidth taken over the small scale
channel fading is
\begin{equation}
\label{equ.E_h_R_i^n}
\mathbb{E}_{\bm h}\{R_i^n\} \approx \log_2(1+\frac{Pr_{ii}^{-\alpha}}{\bar{I_i}}),
\end{equation}
where $\bar{I_i} = P\sum_{j=1,j \neq i}^{B}r_{ij}^{-\alpha} $.
Since channel fading and user location are with independent distribution, the average throughput of a  N-Coop link per unit bandwidth taken over both channel fading and user location
can be obtained as
\begin{equation}
\label{equ.bar_R_i^n}
\begin{split}
&\bar{R_i^n} =\mathbb{E}_{\bm p} \{ \log_2(1+\frac{Pr_{ii}^{-\alpha}}{\bar{I_i}}) \}.
\end{split}
\end{equation}

Because the joint probability density function (PDF) of the distances among D2D users is hard to obtain, we introduce the first order approximation to derive
the expression of $\bar{R_i^n}$. Specifically, for a random variable $X$, the expectation of a function of $X$, $\varphi( X)$, can be approximated as
\cite{approx}
\begin{equation}
\label{equ.1_approx}
\begin{split}
&\mathbb{E}\{\varphi( X)\} = \mathbb{E}\{\varphi(\mu_x +  X - \mu_x )\}\\
& \approx \mathbb{E}\{\varphi(\mu_x) + \varphi'(\mu_x)( X-\mu_x)\} = \varphi(\mu_x),
\end{split}
\end{equation}
where $\mu_x = \mathbb{E}\{ X\}$.

With this approximation, $\bar{R_i^n}$ in (\ref{equ.bar_R_i^n}) is approximated as
\begin{equation}
\label{equ.bar_R_i^n2}
\bar{R_i^n} \approx \log_2(\mathbb{E}_{\bm p}\{Pr_{ii}^{-\alpha}+\bar{I_i}\}) - \log_2(\mathbb{E}_{\bm p}\{\bar{I_i}\}).
\end{equation}
To simplify the analysis, interference generated by each surrounding cluster is regarded statistically identical. Then, the interference link
distance $r_{ij}$ has the same distribution $f(r)$, where $r=r_{ij}/D$. The PDF of the signal link distance $r_{ii}$ can be obtained from \cite{PDF}
by variable substitution $r=r_{ii}/D$ as
\begin{eqnarray}
\label{equ.g_r_ii}
g(r)=\frac{1}{D}
\begin{cases}
2r(r^2-4r+\pi), &0\leq r<1 \\
8r\epsilon-2r(r^2+2)\\
+4r{\arcsin(\frac{1}{r})-\arccos(\frac{1}{r})}, &1\leq r<\sqrt{2}\\
\end{cases}.
\end{eqnarray}

Using similar way as that in  \cite{PDF} and after some tedious derivations, we can obtain the PDF of the interference link
distance $r_{ij}$ as follows,
\begin{eqnarray}
\label{equ.f_r_ij}
f(r)=\frac{1}{D}
\begin{cases}
2r^2-r^3, &0\leq r<1 \\
2r-4r^2+2r^3-2r\epsilon+\frac{2r}{\epsilon}\\
-\frac{2r^3}{\epsilon}+4r\arcsin(\frac{\epsilon}{r}), &1\leq r<\sqrt{2}\\
4r\epsilon+4r\arcsin(\frac{1}{r})\\
-r-4r^2, &\sqrt{2}\leq r<2\\
-5r-r^3+4r\epsilon\\
-4r\arcsin(\frac{\xi}{r})-\arcsin(\frac{1}{r})\\
-\frac{4r}{\xi}+r\xi+\frac{r^3}{\xi}, &2\leq r<\sqrt{5}\\
\end{cases}
\end{eqnarray}
where $\epsilon \triangleq \sqrt{r^2-1}$, and $\xi \triangleq \sqrt{r^2-4}$.

To further simplify the expression, we only consider dominant interference generated from the
nearest eight clusters around the $i$th cluster. Then, from (\ref{equ.bar_R_i^n2}) we can obtain
\begin{equation}
\label{equ.bar_R_i^n3}
\begin{split}
&\bar{R_i^n} \approx \log_2(Q_1(\alpha))-\log_2(Q_2(\alpha))-3,
\end{split}
\end{equation}
where $Q_1(\alpha)\triangleq\int_{0}^{\sqrt{2}}r^{-\alpha}g(r)dr +8\int_{0}^{\sqrt{5}}r^{-\alpha}f(r)dr$, and
$Q_2(\alpha)\triangleq\int_{0}^{\sqrt{5}}r^{-\alpha}f(r)dr$, which are easy to compute numerically with the closed-form expression of
$f(r)$ and $g(r)$.
Note that $\bar{R_i^n}$ only depends on path loss exponent $\alpha$.

\subsubsection{Average Throughput of Coop Link}
In \emph{Mode 1}, all DTs jointly  transmit the requested files to the users with zero-foring beamforming (ZFBF), where the ZFBF can be computed at
BS and broadcasted to all DTs. Then, the SINR of the DR of the active link in the $i$th cluster can be expressed as
\begin{equation}
\label{equ.SINR_R^c}
\gamma_i^c = \frac{P\| \bm h_{i}\|^2\delta_i}{\sigma^2} \approx  \frac{P\sum_{j=1}^{B}r_{ij}^{-\alpha}| h_{ij}|^2}{B\sigma^2} ,
\end{equation}
where $\bm h_{i}=[\sqrt{r_{i1}^{-\alpha}}h_{i1},\sqrt{r_{i2}^{-\alpha}}h_{i2},...,\sqrt{r_{iB}^{-\alpha}}h_{iB}]$ is the composite channel vector
between all DTs and the DR, $0\le \delta_i \le 1$, a larger value of $\delta_i$ indicates a better orthogonality
between $\bm h_{i}$ and $\bm h_{j}$ for $i \ne j$. The approximation comes from the fact $\delta_i \approx (BN^t-B+1)/{B}={1}/{B}$ \cite{ZQ.TVT13}, where $N^t$
is the number of antennas per DT that is one in this paper.

Using the same approximation as deriving (\ref{equ.E_h_R_i^n}), the average throughput of a Coop link per unit bandwidth is obtained as
\begin{equation}
\label{equ.bar_R_i^c}
\begin{split}
&\bar{R_i^c} \approx \mathbb{E}_{\bm p} \{ \log_2(1+\frac{P\sum_{j=1}^{B}r_{ij}^{-\alpha}}{B\sigma^2}) \}.
\end{split}
\end{equation}

Again, by applying the first-order approximation in (\ref{equ.1_approx}), using (\ref{equ.g_r_ii}) and (\ref{equ.f_r_ij}), and only considering
dominant signal, we can further approximate $\bar{R_i^c}$ as
\begin{equation}
\label{equ.bar_R_i^c2}
\bar{R_i^c} \approx \log_2(1+\frac{PD^{-\alpha}}{B\sigma^2}Q_1(\alpha)),
\end{equation}
where $Q_1(\alpha)$ is defined in (\ref{equ.bar_R_i^n3}).
\subsubsection{Average Network Throughput}

Finally, from (\ref{equ.bar_R}), (\ref{equ.bar_R_i^n3}) and (\ref{equ.bar_R_i^c2}) the average network throughput can be obtained as
\begin{equation}
\label{equ.bar_R2}
\begin{split}
\bar{R} &= WBP^c\eta ( \log_2(Q_1(\alpha))-\log_2(Q_2(\alpha))-3)\\
&+WB(1-P^c\eta)\log_2(1+\frac{PD^{-\alpha}}{B\sigma^2}Q_1(\alpha).
\end{split}
\end{equation}
\subsection{Optimizing Bandwidth Partition}
To optimize the value of $\eta$ to maximize the average network throughput while guarantee user fairness, we consider the constraints that the average user throughput is larger than a given value $\mu$.
Since only one D2D link is active in a cluster each time, with
round robin scheduling, from (\ref{equ.bar_R_i^n3}) and (\ref{equ.bar_R_i^c2}) the average Coop and N-Coop \emph{user throughputs} can be
respectively obtained as
\begin{align}
&\bar{R^c_u} \overset{(a)}{\approx} \frac{WB\eta \bar{R_i^c}}{\bar{N^c}} = \frac{WB\eta(\log_2(Q_1(\alpha))-\log_2(Q_2(\alpha))-3) }{\bar{N^c}} \nonumber\\
&\bar{R^n_u} \overset{(b)}{\approx} \frac{WB(1-\eta)\bar{R_i^n}}{\bar{N^n}} = \frac{WB(1-\eta)}{\bar{N^n}}\log_2(1+\frac{PD^{-\alpha}}{B\sigma^2}Q_1(\alpha)), \label{equ.bar_R^c_n}
\end{align}
where $\bar{N^c}=\mathbb{E}\{N^c\}$ and $\bar{N^n}=\mathbb{E}\{N^n\}$ are respectively the average numbers of Coop users and N-Coop users. (a) and (b) come
from the fact that $R_i^c$ and $N^c$ are independent random variables, thus $\mathbb{E}\{R_i^c/N^c\} = \mathbb{E}\{R_i^c\}\mathbb{E}\{1/N^c\} \approx
\mathbb{E}\{R_i^c\}/\mathbb{E}\{N^c\} = \bar{R_i^c}/\bar{N^c} $ according to (\ref{equ.1_approx}).

\subsubsection{Average Numbers of Coop and N-Coop users}
Denote the number of users  in the $i$th cluster who requesting files in $\mathcal{G}_k$ as $n_{ik}$,  $1 \leq k \leq K_0$, $1
\leq i \leq B$, and $ \mathcal{N}_{i} \triangleq \{n_{i1},...,n_{i{K_0}}\}$. Since the users request files independently, the probability that the numbers of users in the $i$th cluster who requesting files in $\mathcal{G}_k$ are $n_{i1},...,n_{i{K_0}}$
can be derived as
\begin{equation}
\label{equ.P_N_i}
\begin{split}
 {p_{\mathcal{N}_{i}}} = \prod_{m=1}^{K_0}C_{K-\sum_{j=1}^{m-1}n_{ij}}^{n_{im}}\prod_{k=1}^{K_0} P_k^{n_{ik}} \overset{(a)}{=} \frac{K!\prod_{k=1}^{K_0} P_k^{n_{ik}}}{\prod_{j=1}^{K_0}n_{ij}!},
\end{split}
\end{equation}
where $(a)$ comes from $C_{n}^{m}C_{n-m}^{k}=\frac{n!}{m!k!(n-m-k)!}$.

Only when all the $B$ clusters hit a file group $\mathcal{G}_k$ and $k \leq K$ (i.e., $n_{ik}>0$  is satisfied for $k \leq K$  and any $i$, $1 \leq i \leq
B$), the users requesting the files within $\mathcal{G}_k$ are Coop users, and we call $\mathcal{G}_k$ a \emph{hit file group}. The number of Coop users for
all hit file groups can be obtained as
\begin{equation}
\label{equ.N_c}
\begin{split}
&{N^c} = \sum_{k=1}^{K}\sum_{i=1}^{B}\zeta(k) n_{ik},
\end{split}
\end{equation}
where $\zeta(k)=\lceil \sum_{i=1}^{B}sgn(n_{ik})-B+1) \rceil ^+$ ($k\leq K$) indicate whether $\mathcal{G}_k$ is a \emph{hit file group}, $sgn(x)=1$ when
$x>0$, otherwise $sgn(x)=0$, and $\lceil \Lambda \rceil ^+= \max(\Lambda,0)$.


Denote $\mathcal{N}=\{\mathcal{N}_1,\mathcal{N}_2,...,\mathcal{N}_B\}$.
$\Phi_\mathcal{N}=\{\mathcal{N}|n_{ik}\geq 0,\sum_{k=1}^{K_0}n_{ik}=K\}$ represents all possible combinations of $\mathcal{N}$. Then, by taking average of $N^c$ in
(\ref{equ.N_c}) over $\Phi_\mathcal{N}$, we can derive the average number of Coop users as
\begin{equation}
\label{equ.bar_N_c}
\begin{split}
&\bar{N^c}
=\sum_{\Phi_\mathcal{N}}\prod_{i=1}^{B}\frac{K!\prod_{k=1}^{K_0} P_k^{n_{ik}}}{\prod_{j=1}^{K_0}n_{ij}!} \sum_{k=1}^{K}\sum_{i=1}^{B}\zeta(k) n_{ik} .
\end{split}
\end{equation}
Considering that $N^n+N^c+N^b=N^n$, we can obtain $\bar{N}^n$ by deriving the average number of cellular users $\bar{N_b}=\mathbb{E}\{N^b\}$.
Since all
requests follow Zipf distribution independently, the number of files that can not be fetched via D2D is a random variable following
Binomial distribution and $N^b \sim B(M,1-\sum_{k=1}^{K}P_k)$. Therefore, $\bar{N^b}=M(1-\sum_{k=1}^{K}P_k)$. Then, the average number of N-Coop users can be derived
as
\begin{equation}
\label{equ.bar_N_n}
\bar{N^n} = M - \bar{N^c} - \bar{N^b} .
\end{equation}

\subsubsection{Optimal Bandwidth Allocation Factor}
The optimal bandwidth partition problem is formulated as follows
\begin{equation}
\label{equ.opt2}
\begin{aligned}
&\max_{\eta}\,\, \bar{R} = WB(P^c\eta \bar{R_i^c}+(1-P^c\eta)\bar{R_i^n})\\
&s.t.\quad
\bar{R}^c_u\geq \mu, ~~ \bar{R}^n_u \geq \mu, ~~ 0< \eta \leq 1
\end{aligned}
\end{equation}
where the expressions of $\bar{R}^c_u$ and $\bar{R}^n_u$ are in (\ref{equ.bar_R^c_n}).

By taking the derivative of $\bar{R}$ in (\ref{equ.opt2}), we have $\frac{\partial \bar{R}}{\partial \bar{\eta}} =
WBP^c(\bar{R_i^c}-\bar{R_i^n})$.
If $\bar{R^c_i} \geq \bar{R^n_i}$, $\bar{R}$ is an increasing function of $\eta$ and the optimal solution is $\eta^*=\frac{WB\bar{R^n_i}-\mu\bar{N^n}}{WB\mu}$, otherwise $\eta^*=  \frac{WB\bar{R^c_i}}{\bar{N^c}\mu}$.
In fact, it is not hard to prove that $\bar{R^c_i} \geq \bar{R^n_i}$
if $\bar{I_i} \geq B\sigma^2$, which is easy to satisfy in D2D communications.
%


Except the system parameters $W$ and $\mu$, the optimal bandwidth allocation factor $\eta^*$ depends on $N_f$, $\beta$, $B$, $M$ and $N$, which can be updated together with the optimal cluster size.

\section{Numerical and Simulation Results}
\label{sec:simulation} In the sequel, we evaluate the performance of Coop strategy via simulation and numerical results. We consider a square hotpot area with the side length $D_c=75$ m, where $M=135$ users are randomly located. Such a setting is the same as \cite{JMY.JSAC}, where $2\sim3$ users are located within every area of $10\times10$ $m^2$.
The
path-loss model is $37.6+36.8\log_{10}(r)$ \cite{JMY.JSAC}. Each user is with transmit power $P=20$ dBm. $W=20$ MHz, and $\sigma^2=-95$ dBm.
The file catalog size $N^f=300$ files, and each user caches $N=20$ files \cite{JMY.JSAC}. The parameter of Zipf distribution $\beta =1$. The user throughput constraint is
 $\mu = 1$ Mbps or $\mu = 2$ Mbps. This setup is used in the sequel unless otherwise specified.

\begin{figure}[!htb]
  \centering
  \includegraphics[width=0.41\textwidth]{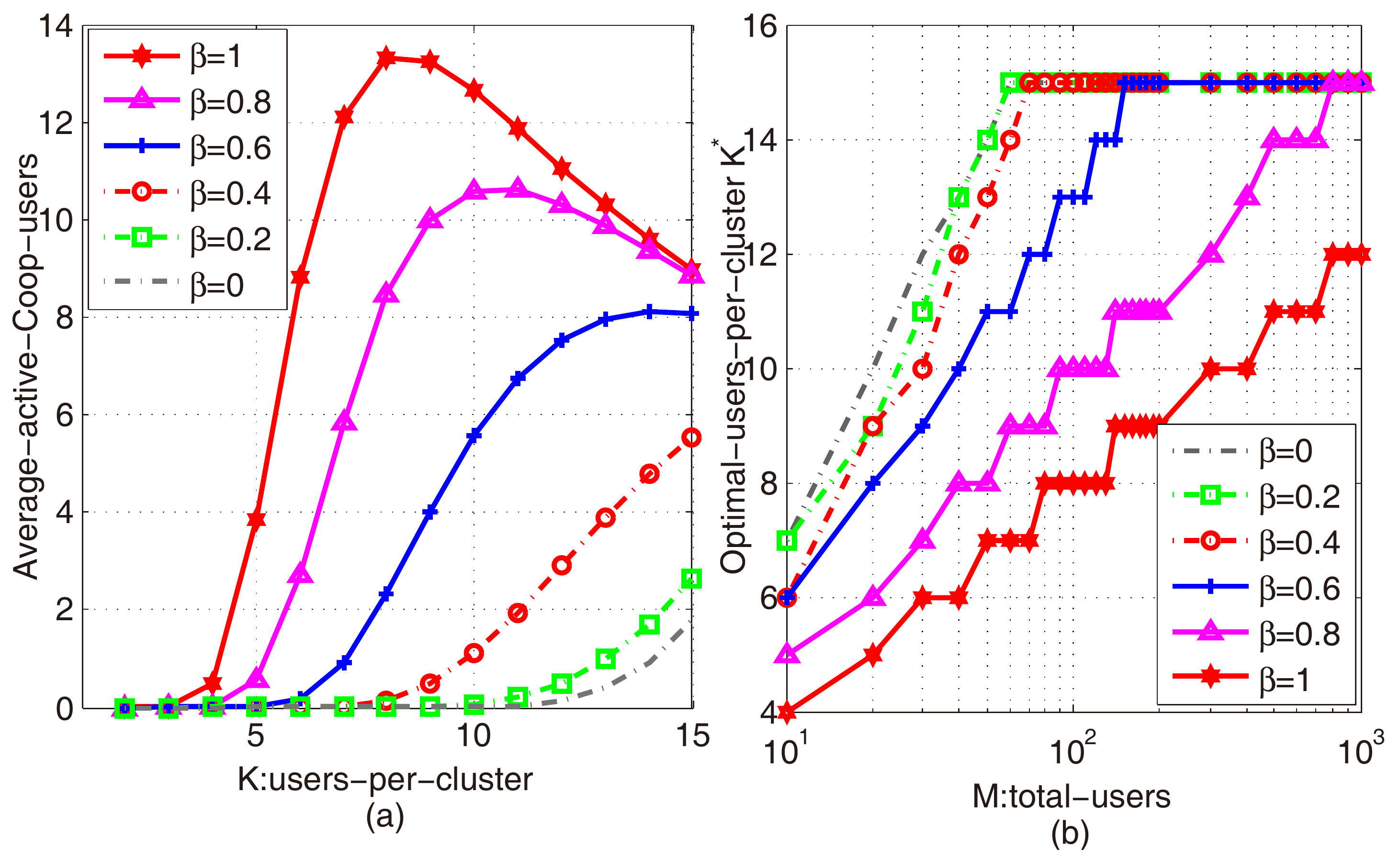}\\
  \caption{$\bar{N^a}$ versus $K$ and optimal cluster size $$$K^*$ versus $M$.}\label{fig.2}
  \vspace{-0.35cm}
\end{figure}

In Fig. \ref{fig.2}, we provide numerical results of the average active Coop users $\bar{N^a}$ for different number of users per cluster $K$ and the optimal cluster size $K^*$ for
different number of users in the hotpot area $M$. As expected, $\bar{N^a}$ increases with $\beta$. By contrast, $K^*$ decreases with $\beta$.
With the growth of $M$, $K^*$ first increases, and then approaches a constant that equals to $K_0=N^f/N$. This is because when $K=K_0$, all files in the catalog have been cached at the users. Assigning more than $K_0$ users to each cluster can not increase the Coop users.

\begin{figure}[!htb]
  \centering
  \includegraphics[width=0.41\textwidth]{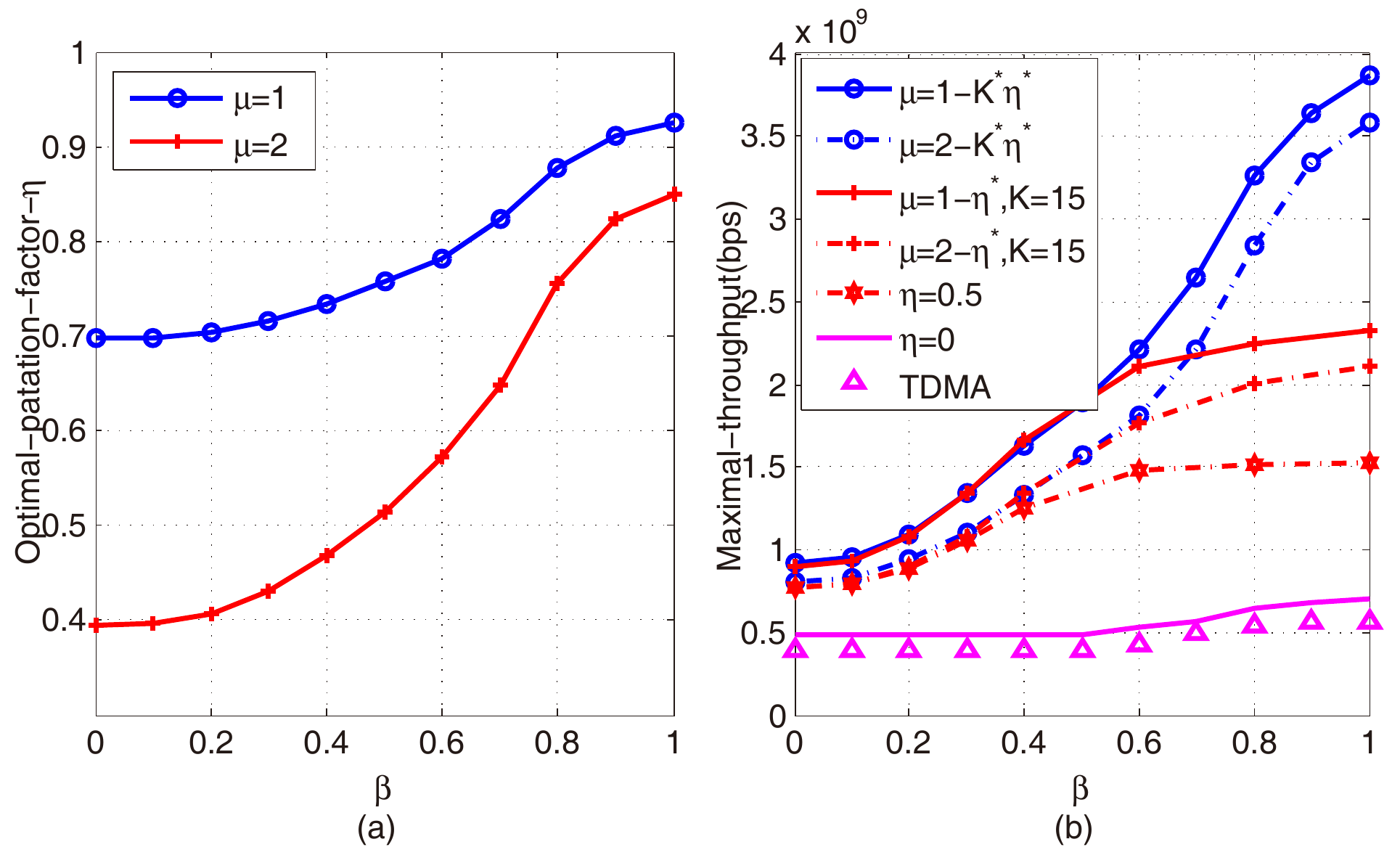}\\
  \caption{$\eta^*$ and maximal average network throughput $\bar{R}$ versus $\beta$. }\label{fig.5}
  \vspace{-0.15cm}
\end{figure}
In Fig. \ref{fig.5}(a), we present the optimal solution of problem  (\ref{equ.opt2}) $\eta^*$ versus $\beta$. As expected, $\eta^*$ increases with $\beta$. However, $\eta^*$
decreases as $\mu$ increases, because more bandwidth is needed for N-Coop users to support higher user throughput.

In Fig. \ref{fig.5}(b), we provide the simulation results for maximal throughput. In the legends, ``$\eta=0$" refers to the strategy in  \cite{Golrezaei.TWC} and ``TDMA" is the strategy in \cite{JMY.JSAC} (where each cluster can select an active D2D link every 4 rounds of scheduling), which serve as the baseline for comparison. ``$\eta=0.5$" refers to a Coop strategy without optimizing $K$ and $\eta$, and ``$\eta^*, K=15$" refers to a Coop strategy without optimizing $K$. We can see that optimizing the cluster size and bandwidth partition become necessary when $\beta >0.4$. With $K^*$ and $\eta^*$,
 the throughput gain over the baseline for $\beta=1$ is $500\% \sim 600\%$. Even when $\beta=0$, the throughput gain is still $40\% \sim 80\%$.

\section{Conclusions}
\label{sec:conclusion} In this paper, we proposed an opportunistic cooperative D2D communication strategy with caching at devices. We optimized the
cluster size and  the bandwidth allocated to Coop and N-coop users. Simulation and numerical
results show that the proposed strategy can improve the throughput significantly, even when the user requests follow uniform distribution.

\bibliographystyle{IEEEtran}
\bibliography{VTC_CBQ_15}

\begin{thebibliography}{10}
\providecommand{\url}[1]{#1}
\csname url@samestyle\endcsname
\providecommand{\newblock}{\relax}
\providecommand{\bibinfo}[2]{#2}
\providecommand{\BIBentrySTDinterwordspacing}{\spaceskip=0pt\relax}
\providecommand{\BIBentryALTinterwordstretchfactor}{4}
\providecommand{\BIBentryALTinterwordspacing}{\spaceskip=\fontdimen2\font plus
\BIBentryALTinterwordstretchfactor\fontdimen3\font minus
  \fontdimen4\font\relax}
\providecommand{\BIBforeignlanguage}[2]{{%
\expandafter\ifx\csname l@#1\endcsname\relax
\typeout{** WARNING: IEEEtran.bst: No hyphenation pattern has been}%
\typeout{** loaded for the language `#1'. Using the pattern for}%
\typeout{** the default language instead.}%
\else
\language=\csname l@#1\endcsname
\fi
#2}}
\providecommand{\BIBdecl}{\relax}
\BIBdecl

\bibitem{Andrews.D2D}
X.~Lin, J.~Andrews, A.~Ghosh, and R.~Ratasuk, ``An overview of {3GPP}
  device-to-device proximity services,'' \emph{IEEE Commun. Mag.}, vol.~52,
  no.~4, pp. 40--48, 2014.

\bibitem{Survey.D2D}
A.~Asadi, Q.~Wang, and V.~Mancuso, ``A survey on device-to-device communication
  in cellular networks,'' \emph{IEEE Commun. Surveys Tuts.}, vol.~16, no.~4,
  pp. 1801--1819, 2014.

\bibitem{Golrezaei.TWC}
N.~Golrezaei, P.~Mansourifard, A.~Molisch, and A.~Dimakis, ``Base-station
  assisted device-to-device communications for high-throughput wireless video
  networks,'' \emph{IEEE Trans. Wireless Commun.}, vol.~13, no.~7, pp.
  3665--3676, 2014.

\bibitem{JMY.JSAC}
M.~Ji, G.~Caire, and A.~Molisch, ``Wireless device-to-device caching networks:
  Basic principles and system performance,'' \emph{IEEE J. Sel. Areas Commun.},
  early access.

\bibitem{IA.D2D}
H.~Elkotby, K.~Elsayed, and M.~Ismail, ``Exploiting interference alignment for
  sum rate enhancement in {D2D}-enabled cellular networks,'' \emph{IEEE WCNC},
  2012.

\bibitem{CD2D.15}
Y.~Cao, T.~Jiang, and C.~Wang, ``Cooperative device-to-device communications in
  cellular networks,'' \emph{IEEE Wireless Commun.}, vol.~22, no.~3, pp.
  124--129, 2015.

\bibitem{GC13.Relay}
C.~Ma, G.~Sun, X.~Tian, K.~Ying, Y.~Hui, and X.~Wang, ``Cooperative relaying
  schemes for device-to-device communication underlaying cellular networks,''
  \emph{IEEE GLOBECOM}, 2013.

\bibitem{C2D.Relay}
Y.~Pei and Y.~chang Liang, ``Resource allocation for device-to-device
  communications overlaying two-way cellular networks,'' \emph{IEEE Trans.
  Wireless Commun.}, vol.~12, no.~7, pp. 3611--3621, 2013.

\bibitem{Lau.Tran13}
A.~Liu and V.~K. Lau, ``Mixed-timescale precoding and cache control in cached
  {MIMO} interference network,'' \emph{IEEE Trans. Signal Process.}, vol.~61,
  no.~24, pp. 6320--6332, 2013.

\bibitem{CSI}
K.~Doppler, C.-H. Yu, C.~Ribeiro, and P.~Janis, ``Mode selection for
  device-to-device communication underlaying an {LTE}-advanced network,''
  \emph{IEEE WCNC}, 2010.

\bibitem{SYN.13}
Y.~Jiang and X.~You, ``Research of synchronization and training sequence design
  for cooperative {D2D} communications underlaying hyper-cellular networks,''
  \emph{IEEE ICC}, 2013.

\bibitem{Zipf99}
L.~Breslau, P.~Cao, L.~Fan, G.~Phillips, and S.~Shenker, ``Web caching and
  {Zipf-like} distributions: Evidence and implications,'' \emph{IEEE INFOCOM},
  1999.

\bibitem{Gamma.2011}
R.~W. Heath, T.~Wu, Y.~H. Kwon, and A.~C. Soong, ``Multiuser {MIMO} in
  distributed antenna systems with out-of-cell interference,'' \emph{IEEE
  Trans. Signal Process.}, vol.~59, no.~10, pp. 4885--4899, 2011.

\bibitem{approx}
A.~Papanicolaou, ``Taylor approximation and the delta method,'' 2009.

\bibitem{PDF}
L.~E. Miller, ``Distribution of link distances in a wireless network,''
  \emph{J. Res. Natl. Inst. Stand. Technol.}, vol. 106, no.~2, pp. 401--412,
  2001.

\bibitem{ZQ.TVT13}
Q.~Zhang and C.~Yang, ``Transmission mode selection for downlink coordinated
  multipoint systems,'' \emph{IEEE Trans. Veh. Technol.}, vol.~62, no.~1, pp.
  465--471, 2013.

\end{thebibliography}
\end{document}